\documentstyle[prd,aps]{revtex}
\newcommand{\beq}{\begin{equation}}
\newcommand{\eeq}{\end{equation}}

\def\be{\begin{equation}}
\def\ee{\end{equation}}
\def\bea{\begin{eqnarray}}
\def\eea{\end{eqnarray}}

\newlength{\dinwidth}
\newlength{\dinmargin}
\begin{document}
\input epsf
\draft
\renewcommand{\topfraction}{0.8}
\twocolumn[\hsize\textwidth\columnwidth\hsize\csname
@twocolumnfalse\endcsname
\preprint{SLAC-PUB-9408, hep-th/0208156, August 21 2002}
\title {\Large\bf Supergravity, Dark Energy and the Fate of the Universe  }
 \author{\bf Renata Kallosh,$^1$ Andrei Linde,$^1$
 Sergey Prokushkin$^1$  and Marina Shmakova$^2$ }
\address{ {$^1$Department
  of Physics, Stanford University, Stanford, CA 94305-4060,
USA}    }
\address{$^2$  SLAC,
 Stanford University, Stanford, CA 94309  }
\maketitle
\begin{abstract}

We propose a description of dark energy and acceleration of the universe in  extended supergravities with de Sitter (dS) solutions. Some of them are related to  M-theory with non-compact internal  spaces. Masses of ultra-light scalars in these models are 
quantized in units of the Hubble constant: $m^2 = n\, H^2$. If dS solution corresponds to a minimum of the effective potential, the universe eventually becomes dS space. If dS solution corresponds to a maximum or a saddle point, which is the case in all known models based on $N=8$ supergravity, the flat universe  eventually stops accelerating and collapses to a singularity. We show that in these models, as well as in the simplest models of dark energy based on  $N=1$ supergravity, {\it the  typical time remaining  before the global collapse  is comparable to the present age of the universe}, $t = O(10^{10})$ years. We discuss the possibility of distinguishing between various models and finding our destiny using cosmological observations.

\end{abstract}
 \vskip2pc]

\tableofcontents{}

 \section{Introduction}
 
\subsection{Dark energy and our future}

Recent observations of Type Ia supernovae and the CMB show that the Universe is accelerating, and it is spatially flat ($\Omega_{\rm tot}= \Omega_{M}+\Omega_D = 1$). Approximately $0.3$ of the total energy density of the universe $\rho_0 \sim 10^{-120} M_p^4 \sim 10^{-29}$ g/cm$^{3}$ consists of ordinary  matter ($\Omega_{M} \approx 0.3$), and  $0.7$ of the energy density corresponds to dark energy ($\Omega_D \approx 0.7$), see \cite{supernova,Bond} and references therein.

One can interpret  the dark energy $\rho_D$ either as the vacuum energy (cosmological constant) $\Lambda \sim 0.7\rho_0$, or as the  slowly changing energy of a scalar field $\phi$ with a vacuum-like equation of state $p_D = w\, \rho_D$,\, $w  \approx -1$.  In either case our universe is supposed to be accelerating and rapidly approaching de Sitter (dS) regime. Therefore it would be tempting to conclude that our universe in the future is going to expand exponentially for an indefinitely long time,  $a\sim e^{Ht}$, even if it is closed.

 However, dS regime may be transient, and the future of the universe may be quite different. For example, in most of the models of dark energy it is assumed that the cosmological constant is equal to zero, and the potential energy $V(\phi)$ of the scalar field  driving the present stage of acceleration, slowly decreases and eventually vanishes as the field rolls to $\phi =\infty$, see e.g.  \cite{Dolgov:gh,Wett,Ratra:1987rm,Armendariz-Picon:2000dh,Albrecht:2001xp}. In this case, after a transient dS-like stage, the speed of expansion of the universe decreases, and the universe reaches  Minkowski regime.
In both cases (dS space and Minkowski space) life may survive for an extremely long time until all protons decay and the energy resources of the nearby part of the universe are depleted.

However, there is another possibility, which for a long time did not attract much attention. It is quite possible that $V(\phi)$ has a minimum at $V(\phi) <0$, or that it does not have any minimum at all and the field $\phi$ is free to fall to $V(\phi) = -\infty$. In this case {\it the universe eventually collapses,  even if it is flat} \cite{MTW,Banks:1995dt,Krauss:1999br,Kaloper:1999tt,Kallosh:2001du,Kallosh:2001gr,Linde:2001ae,Steinhardt:2001st,Felder:2002jk,Heard:2002dr}.

Thus, depending on the choice of the model describing dark energy, the flat  universe with $\Omega=1$ may eventually become dS space, or Minkowski space, or collapse. (It can never become AdS space with energy density dominated by  a negative cosmological constant \cite{Kallosh:2001gr,Linde:2001ae,Felder:2002jk}.)

 \subsection{Dark energy in extended supergravity}

The main goal of this paper is to investigate the possibility to describe dS space and dark energy in supergravity. 
 We will mainly concentrate on the extended gauged supergravities, although we will also discuss some models based on $N=1$ supergravity.
The extended supergravities are most interesting for two reasons. First of all, they  have much closer relation to M/string theory than the ordinary N=1 supergravity. In particular, the maximal $d=4$ $N=8$ supergravity has the same amount of supersymmetries as  M-theory, $8\times 4=32$ (there are   32  supersymmetries in $d=11$ M-theory). Also,   in the context of theories with extra  dimensions the standard N=1 supergravity (with 4 supercharges) simply does not exist in the bulk; the smallest possible supersymmetry in $d=5$ has 8 supercharges, $2\times 4=8$ and therefore $N\geq 2$ in $d\geq 5$.

In $N=1$ supergravity it is  easy to  construct a nonvanishing scalar potential (the $F$-term potential) by choosing a proper superpotential $W$. The freedom of choice is almost unlimited, which does not provide us with strict guiding principles that could help us to construct a realistic theory. Meanwhile, in the extended supergravity one does not have this freedom. Nonvanishing potentials appear only after  gauging of some of the global symmetries of the theory. In a certain sense, the total potential  is similar to the $D$-term potential in $N=1$ theory: it disappears for vanishing gauge coupling $g$. 

Recently it was found that one can describe dark energy in some $d=4$ extended gauged supergravities     that have  dS solutions \cite{Kallosh:2001gr,Kallosh:2002wj}.  Some of these supergravities solve the equations of motion of M-theory with non-compact internal spaces \cite{Hull:1988jw}.

These dS solutions correspond to the extrema of the effective potentials $V(\phi)$ for some scalar fields $\phi$. An interesting and very unusual feature of  these scalars in all known theories with $N\geq 2$ is that their mass squared is quantized in units of the Hubble constant $H_0$ corresponding to dS solutions:
${m^2\over H_0^2}=n $,
where $n$ are some integers of the order 1. 
This property was first observed in \cite{Kallosh:2001gr} for a large class of extended supergravities with unstable dS vacua, and confirmed and discussed in detail more recently  in \cite{Kallosh:2002wj} with respect to a new class of $N=2$ gauged supergravities with stable dS vacua \cite{Fre:2002pd}. The universality of the relation ${m^2\over H_0^2}=n $ may be attributed to the fact that  ${m^2\over H_0^2}$ is an eigenvalue of the  Casimir operator of  dS group,  \cite{Kallosh:2002wj}.

The meaning of this result can be explained in the following way. Usually  the effective potential near its extremum can be represented as $V(\phi) = \Lambda + m^2\phi^2/2$, where $\Lambda$ and $m^2$ are two free independent parameters. However, in extended supergravities with $\Lambda > 0$ one always has $m^2 = n \Lambda/3$, where $n$ are integers (we are using units $M_p = 1$). Taking into account that in dS space $H_0^2 = \Lambda/3$, one has, for $|\phi| \ll 1$, $ 
V(\phi) = \Lambda(1 + n \phi^2/6) = 3H^2_0(1 + n \phi^2/6)$.
In particular, in all known versions of $N=8$ supergravity dS vacuum corresponds to an unstable maximum, $m^2 = -6H_0^2$ \cite{Kallosh:2001gr,Kallosh:2002wj}, {\it i.e.} at $|\phi| \ll 1$ one has
\be \label{simplepot}
V(\phi) = \Lambda(1 -\phi^2) = 3H^2_0(1 -\phi^2) \ .
\ee
Meanwhile, for the $N=2$ gauged supergravity with stable dS vacuum found in \cite{Fre:2002pd} one has $m^2 = 6H_0^2$ for one of the scalars,  and for $|\phi| \ll 1$ one has
\be
V(\phi) = \Lambda(1 +\phi^2) = 3H^2_0(1 +\phi^2) \ .
\ee

 In \cite{Kallosh:2001gr,Linde:2001ae} it was explained that the `fast-roll' regime with $ {m^2\over H^2_0}=O(1)$ is suitable for the late inflation describing the present stage of acceleration of the universe. If one takes  $\Lambda \sim \rho_0 \sim 10^{-120}$  in units $M_p = 1$ and $H_0 \sim 10^{-33}$ eV,  corresponding to the present stage of expansion of the universe, then the quantization rule implies that there are ultra-light scalars with the mass of the order
$
|m ^2|\sim H^2_0 \sim (10^{-33}eV)^2.
$
Here one should distinguish between the time-dependent Hubble constant $H^2(t) = (\rho_M + \rho_D)/3$ and its value $H_0^2 = \Lambda/3$ in  dS regime where $\rho_M=0$ and the dark energy field $\phi$ stays at the minimum/maximun of its potential. However, at the present time, with $\Omega_D \approx 0.7$, one has $H(t) = O(H_0)$.

In the early universe the ultra-light scalar fields may stay away from the extrema of their potentials; typically they `sit and wait' and start moving only when the Hubble constant $H(t)$ determined by cold dark matter, decreases and becomes comparable to $|m|$. We will see  that this could result in noticeable changes of the effective cosmological constant during the last 10 billion years. The existence of this effect can be verified by  observational studies of the acceleration of the universe.

Extended supergravity may lead not only to  potentials with dS extrema, but also to  exponential potentials.
We will show that a particularly interesting potential  $V \sim e^{\sqrt 2\phi}$ can be derived in several models based on extended supergravity.  Despite the current lore \cite{Albrecht:2001xp}, the theory with this potential can describe dark energy and the present  stage of acceleration of the universe, and it does not suffer from the problems with the existence of the event horizon discussed in  \cite{Hellerman:2001yi}.

An important feature of extended gauged supergravities is the fact that {\it quantum corrections to the cosmological constant as well as to the ultra-light  masses} are related to the  value  of the  cosmological constant $\Lambda$ defining the scale of SUSY breaking. For $\Lambda \sim 10^{-120} M_p^4$  these quantum corrections are very small. Thus extended supergravities provide an example of a model where ultra-light scalars $|m^2| \sim  \Lambda$ naturally appear  and are protected against quantum corrections.

\subsection{Supergravity and the fate of the universe}

The relation $|m^2| \sim H^2$  appears not only in extended
supergravity. It is  rather common in $N=1$ supergravity for the
moduli fields that have vanishing mass at $M_p\rightarrow \infty$
\cite{Dine:1983ys}. In contrast to the  extended supergravities,
in the $N=1$ case it is possible to avoid this relation: one can
take a non-minimal K\"{a}hler potential, fine-tune the
superpotential, and/or introduce non-trivial D-terms, what will
modify  the mass/Hubble  ratio in a significant way.  Indeed, in
the supersymmetry breaking hidden sector one should avoid the
relation $|m^2| \sim H_0^2$, in order to avoid huge cosmological
constant $ \Lambda > (10^3\, {\rm GeV})^4$. However, there is no
need to make this fine-tuning in the dark energy hidden
sector\footnote{A hidden sector for quintessence models in $N=1$
supergravity, different from the supersymmetry breaking hidden
sector, was proposed in \cite{Binetruy:1998rz,Brax:2001ah}.}.

In particular, the relation $|m^2| \sim H^2$ occurs in the
so-called supergravity  quintessence model  with dS minimum
\cite{Brax:2001ah}. Although the potential of a truncated model in
this case is simple,  the supergravity model \cite{Brax:2001ah} is
rather complicated, involving many fields with non-minimal
K\"{a}hler potentials and a set of additional assumptions. In
order to study a much simpler toy model for dark energy in $N=1$
supergravity we will consider a Pol\'{o}nyi-type model
\cite{Polonyi:1977pj} of the dark energy hidden  sector.  It has a
minimal K\"{a}hler potential and the simplest superpotential $W(z)
= \mu^2 (z+\beta)$. The parameter $\mu$ should be taken extremely
small,  $\mu^4 \sim \rho_0 \sim 10^{-120} M_p^4$. In this case
there is no need to make the standard fine-tuning  $\beta =
2-\sqrt 3$ in order to avoid the large cosmological constant. For
$|\beta| < 2-\sqrt 3$ the potential has dS minimum with $\Lambda
\sim +10^{-120} M_p^4$. For larger values of $\beta = O(1)$ the
potential has a minimum with $\Lambda \sim -10^{-120} M_p^4$. As
we will see, in both cases this model, just like the models based
on $N=8$ and $N=2$ supergravity, can describe the present state of
acceleration of the universe. However, the  future of the
universe does depend on the choice of the model.

Another interesting model  is the axion quintessence  \cite{Frieman:1995pm,Choi:1999xn}. In the M-theory motivated version of this model proposed in \cite{Choi:1999xn} one has $V(\phi) \sim \Lambda \cos(\phi/f)$; for $f = O(M_p)$ one finds $m^2 = V''(0) = -O(H^2_0)$. As we will show, this version of the axion quintessence model can successfully describe the stage of acceleration of the universe, but, just like the $N=8$ models, it leads to a global collapse of the universe in the future.

In order to obtain a fully realistic model of dark energy, one
would need to construct theories involving the observable sector,
the hidden sector responsible for supersymmetry breaking, and the
hidden dark energy sector. This is a complicated and as yet
unsolved problem. The supersymmetry breaking scale in the  dark
energy hidden  sector is of the order $ M_{susy}^{\rm D }\sim
10^{-12}\, \rm GeV \ $. Meanwhile, in the supersymmetry breaking
hidden sector the corresponding scale  is $M_{susy}\sim 10^{3} \,
\rm GeV$ or even much greater. This means that  the difference
between these two types of  supersymmetry  breaking  is more than
$15$ orders of magnitude. Even though the fields from each of
these sectors may interact with each other only gravitationally,
this interaction may be strong enough to alter the important
relation $m = O(H_0)$ for the ultra-light scalars. This problem
and various ways to address it were discussed in
\cite{Binetruy:1998rz,Brax:2001ah,Kolda:1998wq,Choi:1999xn}.

In this respect, extended supergravity may be particularly
interesting as  the dark energy hidden sector if the mysterious
mass quantization rule $m^2 =  n\,H_0^2$ has some fundamental
meaning and remains stable with respect to the interaction of the
ultra-light scalars with the fields from the observable sector.
One may even argue that the reason for using extended $N\geq 2$
supergravities is due to the nature of gravitational and vector
fields that may live in five dimensions or higher, where $N=2$ is
the smallest supersymmetry available. However,  realistic models
of supersymmetry breaking in the context of supergravity, branes
and extra dimensions are  yet to be developed.

For the time being, one may consider the simple models of dark
energy based on supergravity as toy models with some interesting
and very unusual  features that could be studied by cosmological
observations.

In particular, if dS vacuum corresponds to a minimum of the
effective potential, as in the $N=2$ model of Ref.
\cite{Fre:2002pd} and in the Pol\'{o}nyi model with $|\beta| <
2-\sqrt 3$, the universe asymptotically approaches a stable dS
regime.

On the other hand, if the effective potential is negative at the
minimum, $V(\phi_0) <0$, as in the Pol\'{o}nyi model with $|\beta|
> 2-\sqrt 3$ and in the axion quintessence model with  $V(\phi) \sim \Lambda \cos(\phi/f)$, or if it is unbounded from below, as in all known
versions of $N=8$ supergravity admitting dS solutions, the flat
$\Omega=1$ universe eventually  collapses.  The typical time
remaining before the collapse in these models is $O(m^{-1})\sim
H_0^{-1} $. Since the total age of the universe now is also given
by $O(H_0^{-1})$, in this class of models {\it  the time remaining
before the global collapse  is of the same order as the present
age of the universe,} $t_{{}_{\rm Big Crunch}} \sim 10^{10}$
years.

Thus, in this paper we will study the supergravity  models  which
are able to describe nicely the present and the past, and may
predict  an  ultimate collapse  that may occur within the next
10-20 billion years. We will show that some of the models
predicting gravitational collapse within the next 5 billion years
can be ruled out by observational data. The investigation of the
stability of the universe on a greater time scale will require a
much more detailed observational study of the present stage of
acceleration of the universe.

\section{Supergravities with de Sitter Solutions and Dark Energy }
In this paper we will consider late stages of evolution of  universe, as predicted by inflationary theory. This means that the observable part of the universe is flat and nearly homogeneous, with  the metric
\bea
ds^2= dt^2- a^2(t)\ d\vec x^2 \  .
\eea
We will assume that in addition to cold dark matter with energy density $\rho_M$ the universe also contains a slowly changing  field $\phi$. This field may take different values in different parts of the universe, but inside our horizon the field is almost exactly homogeneous.

Another assumption is that this field has an extremely flat effective potential $V(\phi)$, and the energy density of the post-inflationary universe was dominated not by this field, but by radiation and, later on, by cold dark matter with density  $\rho_M(t) =  \rho_M(t_0)\left({a(t_0)\over a(t)}\right)^3$. Therefore even if the field $\phi$ rapidly moved in the early universe, eventually it lost its speed due to the effective `friction' related to the  redshift  of its kinetic energy in an expanding universe. Since that time the field sits and waits at some point $\phi_0$ until the expansion of the universe slows down and the field becomes free to move again.

The initial position  $\phi_0$  of the field at this stage is determined by the processes in the early universe. If one assumes that $V(\phi)$ is extremely flat, and that it was flat during inflation, then $\phi_0$ can take all possible values in different parts of the universe, with probability that does not depend on $\phi_0$, see e.g. \cite{LLM,Malquarti:2002bh}. It is also possible that in the early universe the effective potential was quite different. It could have a deep minimum at some particular value  $\phi_0$, and then the position of the minimum could change when the density of matter in the universe becomes sufficiently small \cite{Dine:1983ys}. For simplicity, in this paper we will assume that this is not the case and that $V(\phi)$ was always flat. However, this is an important and interesting issue to be studied in the future.

Let us consider for a moment an important case where $V(\phi) = \Lambda = const$, or assume that $V(\phi)$ does depend on $\phi$ but $\phi_0$ corresponds to an extremum of the effective potential, $V'(\phi_0) = 0$. In this case (if one ignores quantum fluctuations) the scalar field remains at the same point $\phi_0$ even after the density of cold dark matter becomes small. Gradually the density of the universe becomes dominated by $V(\phi_0) = \Lambda$.

At the stage when the universe was cold dark matter dominated, its scale factor obeyed the simple equation $a(t) \sim t^{2/3}$, and the Hubble constant $\dot a/a$ was given by  
$H(t) = {2\over 3 t}$. 
Once the universe becomes dominated by $V(\phi_0) = \Lambda$, this regime switches to $a(t)= e^{H_0t}$, where $H_0 = \sqrt{\Lambda/3}$ (in units $M_p=1$). This is the regime of late inflation that can be described in terms of dS space.

De Sitter space is defined as a hypersurface in a 5d space
$-T^2+{1\over c^2}(X^2+Y^2+Z^2+W^2)=H_0^{-2}.$
Here  $c$ is the speed of light (later we will use the units  $c = 1$).  In terms of  length scale we have
\be
 -X_0^2+(X^2+Y^2+Z^2+W^2) = R_0^2  \ .
\label{ds} \ee
Here $R_0=c H_0$ is the event horizon of dS universe: an observer in an exponentially expanding universe $a(t)\sim e^{H_0 t}$ sees only those events that take place at a distance no farther away than $R_0$.   

Note that this regime begins at $\rho_M \sim \Lambda$. Until that moment one can use the relation $H = {2\over 3 t}$. After that $H$ decreases a bit and becomes constant, $H = H_0$. This implies that the switch between the two regimes occurs at the cosmological time $t \sim 0.5 H_0^{-1}$. Numerical investigation shows that the present moment, when the total density of cold dark matter is about $0.3 \rho_0$ and $\Lambda$ is about $0.7 \rho_0$, corresponds to the time $t \approx 0.8 H_0^{-1}$ after the end of inflation. In this regime $H_0 = \sqrt{\Lambda/3} \approx \sqrt {0.7\rho_0/3}\approx \sqrt {0.7} H$, where $H$ is the present value of the Hubble constant. Therefore one can re-express the present age of the universe in terms of $H$: \ $t \approx 0.8 H_0^{-1} \approx 0.96 H^{-1}$. The last number will often appear in our figures since we will measure time in units of the present value of $H^{-1}$.

The best estimate of the present age of the universe is between 13.6 to 14 billion years \cite{Bond}. This yields the value $H_0^{-1} \sim 11$ billion years, which corresponds to the size of the event horizon in dS space $R_0 =cH_0^{-1} \approx 10^{28}$ cm. This corresponds to $10^{61} l_{p}$  which is the minimal size of  dS hyperboloid, the size of its throat. Here $l_{p}\sim 10^{-33}cm $.
The corresponding scale of energies is $10^{-61} M_{p}\sim 10^{-33}$ eV.

Now let us see whether this cosmological picture  can be related to supergravity.
A generic four-dimensional gauged supergravity which will be used for dark energy hidden sector has  part which includes gravity coupled to $2n$ scalar fields $\phi^i$ and a potential
$$
g^{-1/2} L = -{1\over 2} R +  {1\over 2}\,G_{ij}(\phi)\,  \partial_\mu \phi^i \partial_\nu \phi^j\, g^{\mu\nu} -V(\phi) \ .
$$
Here $G_{ij}(\phi)$ is the metric in the moduli space and we have not included vector fields and fermions.

As explained  in  \cite{Kallosh:2002wj}, in all known  gauged extended supergravities possessing dS solutions,  the eigenvalues $V''$ of the mass matrix $(m^2)^i_{j}$ in dS extremum are proportional to the value of the potential $ V $ in the extremum. This corresponds to integer values of the ratio of eigenvalues of the mass matrix to $H^2_0$ in the range between $12$ and $-6$. The mass matrix at the  dS critical point is defined as follows:
\be
(m^2)^i{}_{j} =  G^{ik}(\phi) \partial_k \partial _j V|_{\phi_{cr}} \ .
\ee
Here $\phi^i_{cr}$ are the critical values of the scalar fields corresponding to the extrema of $V$.

We assume that there is a dark energy Lagrangian given by extended supergravity with dS solution and that in addition to the scalars  $\phi^i$ representing the dark energy of the universe   there is also the usual cold dark matter energy density $\rho_M$ contribution on the right hand side  of the Friedmann equations:

\be
\ddot \phi^i + 3{\dot a\over a} \dot \phi^i + \Gamma^i_{jk}\dot \phi^j \dot \phi^k  + G^{ij}  {\partial V\over \partial \phi^j} =0 \ ,
\ee

\be
H(t)\equiv  {\dot a\over a}= \sqrt{{\rho_M + V + E_{kin}\over 3}}\ .
\label{F1}\ee
Here $a(t)$ is the scale factor of the  flat FRW metric,   $\rho_M= {C\over a^3}$.
The kinetic part of the dark energy, $E_{kin}(\phi)$, is given by
$
E_{kin}=   {1\over  2 } G_{ij}(\phi)  \dot \phi^i \dot \phi^j  
$ and $\Gamma^i_{jk}(\phi)$ is the Christoffel symbol in the moduli space defined by the metric $G_{ij}(\phi)$.

In the linear approximation for the fields $\phi$ which are close to their critical values $\phi_{cr}$ (i.e. to the extrema of $V(\phi)$) the scalar field equations are reduced to
$
\ddot {\delta \phi^i} + 3 H \dot {\delta \phi^i}   +( m^2)^i{}_j  \delta \phi^j =0 $
where $\delta \phi^i= \phi^i- \phi^i_{cr}$. This shows the  universal character of  dS critical points in all known gauged supergravity models: all scalar equations near dS critical point take the form
\be
\ddot {\delta \phi^i} + 3 H_0 \dot {\delta \phi^i}   + n H^2_0  \delta \phi^i =0 \ ,
\ee
where $n= {m^2\over H^2_0}$, the Casimir operator of  dS group,  takes values from $-6$ up to $+12$.

The energy density $\rho_D$ and the pressure $P_D$ for the supergravity dark energy are given by
\bea
\rho_{D}&=& E_{kin}+ V = {1\over  2 } G_{ij}(\phi) \dot \phi^i \dot \phi^j  + V(\phi) \ ,\\
P_{D}&=& E_{kin}- V = {1\over  2 } G_{ij}(\phi) \dot \phi^i \dot \phi^j  - V(\phi) \ .
 \eea
 The total energy includes also the energy of matter
 \be
\rho_{T}= \rho_M+ E_{kin}+ V \ .
 \ee
A dark energy (matter) density  is given by a ratio of the dark energy (matter) to the total energy.
\be
\Omega_{D}= {\rho_D\over \rho_{T}} \ , \qquad \Omega_{M}= {\rho_M\over \rho_{T}} \ .
\ee
Experimentally now $\Omega_M \sim 0.3$ and $\Omega_D \sim 0.7$ and  $\Omega_T =\Omega_M  + \Omega_D   \sim 1$.
Thus when we plot the time dependence of $\Omega_{D}$ the  present time will be specified by  $\Omega_D = 0.7$.
Another important characteristics of the dark energy is its pressure-to-energy ratio defining dark energy equation of state $P_D= w_D \rho_D$:
\be
w_D= {P_D \over \rho_D}= {E_{kin}- V\over E_{kin}+ V} \ .
\ee
In what follows we will skip the index $D$ in $w_D$ and use  $w$ for dark energy  $ w_D$, since we consider only recent period when radiation is irrelevant and $p_M=w_M=0$.

The parameter  $w$ in general strongly depends on time $t$. It is convenient to introduce an average value of $w$ which is useful for understanding an integrated influence of $w$ on observations:
\be \label{barw}
\bar w = {\int da\ \Omega_D(a)\ w(a)\over  \int da\ \Omega_D(a)}   \ .
\ee

Note that the deceleration parameter (by the time when radiation is negligible) is given by
\begin{eqnarray}
q(t)\equiv -{\ddot a\over a H^2}&=& {1\over 2} [\Omega_M+ \Omega_D(1+3 w)] \nonumber\\&\sim & {1\over 2} + {3\over 2}\Omega_D w \ .
\label{decel}\end{eqnarray}
Here we used the fact that $w_M=0$.
At present $q_0\sim  {1\over 2}+ w$ is close to $-0.5$ and therefore $w$ is close to $-1$.

\subsection{Collapsing universe}

 In all models with  dS maximum or saddle point the universe will eventually collapse, i.e. the scale factor will shrink to zero.
 At some point in time $\tilde t$  the potential becomes negative and cancels the positive contribution of $\rho_M$ and $E_{kin}$ in   the expression for the  Hubble parameter:
\be
H^2(\tilde t)= {\rho_M + V + E_{kin}\over 3}=0 \ .
\ee
At this moment the universe stops expanding; the  Hubble parameter  becomes negative for $t>\tilde t$,  which inevitably leads to a collapse \cite{MTW,Banks:1995dt,Krauss:1999br,Kaloper:1999tt,Kallosh:2001du,Kallosh:2001gr,Linde:2001ae,Steinhardt:2001st,Felder:2002jk,Heard:2002dr}.
In case of  collapsing universe, we have to use an alternative form of one of the Friedmann equations.  Instead of Eq. (\ref{F1}) we will use
\be
  {\ddot a\over a}= {-\rho_M + V -2 E_{kin}\over 3}
\label{F2}\ee
since Eq. (\ref{F1}) is not well defined in the case where $H$ vanishes and becomes negative.
The equation of state function $w(t)$ will grow from the initial value $-1$ to infinity somewhere in the future, well before the collapse, and then, near the singularity, it usually approaches the value $+1$ corresponding to the stiff equation of state $p = \rho$ \cite{Felder:2002jk,Heard:2002dr}.
$$
w^{initial}=-1, \quad w^{today}= -1+\delta_{col}, \quad w^{future}\rightarrow +1 .$$

\subsection{Future dS universe}

In all models  based on extended supergravities with dS minimum of the potential, $w$ will start at $-1$ at the time when all ultra light fields are frozen somewhere at the slope of the potential, so that $E_{kin}^{initial}=0$. By the time when the Hubble parameter decreases to the value comparable with the mass of these fields, they will start moving towards the minimum of the potential with some velocity. This will raise the value of $w $ above $-1$. However, eventually all fields will approach the minimum and stabilize at $E_{kin}^{attractor}=0$, perhaps after some oscillations. Again, the attractor value $w=-1$ will be reached in the long term future. Depending on initial conditions of the fields, 
 at present  $\dot w$ may be positive, or negative or vanishing and $w$ takes  any value above $-1$.
This would mean that  during the last Hubble time it could be increasing monotonically or first increasing and later decreasing.
$$
w^{initial}=-1, \quad w^{today}= -1+\delta_{attr}, \quad w^{future}=-1.
$$

\

For  general supergravity models, we find numerical solutions of the Friedmann equations for the universe evolution during the last Hubble time. We pay  particular attention to time dependence of   $\Omega_D(t)$ and $w(t)$  since they can help to confirm or rule out the class of dark energy models with future dS space or collapse. All models are adjusted so that today corresponds to $t_0=0$, $t$ is negative in the past and positive in the future. We  use units in which $H^2=1$ at present and assume that $\Omega_D=0.7,\ \Omega_M=0.3$ today.

For each model we  also present the evolution as a function of the redshift $z$ where ${a(t_0)\over a(t)}=1+z$ and $z_{today}=0$. We  only consider non-negative $z\geq 0$, which correspond to the past and present since only such $z$ are subject to cosmological observation.

\section{Gauged Supergravities and Collapsing Universe}

\subsection{$N=8$  supergravity and M-theory}
Our particular interest in the cosmological aspects of the theory with maximal supersymmetry is due to the common trend to have maximal amount of supersymmetry allowed by experimental and theoretical considerations.

M-theory supergravity or supergravities related to low energy string theory, compactified on spaces other than torus, may lead to non-trivial  potentials. However, if the compactification volume is finite the relevant four dimensional supergravities do not have  dS solutions \cite{Gibbons:85}.

Still some four-dimensional gauged  $N=8$ supergravities with 32 supersymmetries have dS solutions, and some of them are directly related to M/string theory \cite{Hull:1988jw}. In agreement with the no-go theorems of \cite{Gibbons:85} they correspond to solutions of M/string-theory with non-compact internal  seven (six) dimensional space.\footnote{There exists  only one known version of quintessence derived from M-theory \cite{Townsend:2001ea}. It starts with $SO(3,3)$ gauged supergravity in five dimensions which has dS solution, related to eleven-dimensional supergravity with non-compact six-dimensional space. The five-dimensional dS solution in this model is unstable \cite{Townsend:2001ea,Hull:2002cv}, so it is likely to lead to a  universe collapsing in the future, just as all $N=8$ models considered in this section.}

 Recently the meaning of such M/string theories  with non-compact internal space was reconsidered \cite{Gibbons:2001wy}. The relation between  states of the higher dimensional and the four dimensional theory in such backgrounds is complicated since the standard Kaluza-Klein procedure is no longer valid. It is nevertheless true that the class of four-dimensional dS supergravities which we will consider below as dark energy candidates has a direct link to M/string theory. They solve equations of motion of the eleven-dimensional supergravity with 32 supersymmetries.  Moreover, these theories are perfectly consistent from the point of view of the four-dimensional theory: all kinetic terms for scalars and vectors are positive definite.

One of the simplest solutions of 11d supergravity is given by a warped product space of the four-dimensional dS space and the seven-dimensional hyperboloid $H^{p,q}$. Our fiducial model with all scalars constant is defined by the dS surface (\ref{ds}), and by the internal space $H^{p,q}$
given by the equation $\eta_{AB} z^A z^B= \alpha \, R_0^2 $ in an eight-dimensional space. Here the constant $\alpha$ depends on $p,q$. The metric $\eta_{AB}$ is constant and has $p$ positive eigenvalues and $q$ negative eigenvalues,  $p+q=8$.

The simplest (and typical) representative of dS supergravities  originated in M-theory, has the potential \cite{Hull:rt} with dS maximum:
\be
 V= \Lambda (2-  \cosh {\sqrt 2} \phi)
 \label{Hull}\ee
(for canonically normalized fields $\phi$).

In this model the extremum of the potential is at $\phi = 0$. $V(0)= \Lambda$, $V''(0) = -2 V(0)$, which corresponds to $m^2=-6H^2_0$.
This is the four-dimensional $N=8$ supergravity with $p=q=4$, which has gauged $SO(4,4)$ non-compact group. At the dS vacuum it is broken down to its compact subgroup, $SO(4)\times SO(4)$. The value of the cosmological constant $\Lambda$ here is related to the value of the gauge coupling $g$ as follows: 
$$
\Lambda = 2 g^2 \ .
$$
Near the extremum the potential can be represented by the simple quadratic expression $V(\phi) = \Lambda(1 -\phi^2)$, see Eq. (\ref{simplepot}). It is unbounded from below, so the theory is unstable, and the first idea would be to discard such models altogether. However, the potential
 remains positive for $|\phi| \lesssim 1$, and for small $\Lambda$ the time for the instability to develop  can be much greater than the present age of the universe, which is quite sufficient for our purposes \cite{Kallosh:2001gr}. In fact, we will see that this instability allows us to avoid the standard fine-tuning problem plaguing most of the versions of the theory of quintessence \cite{KallLin}.

The gauge coupling as well as the cosmological constant in four-dimensional supergravity have the same origin in M-theory: they come from the flux  of an antisymmetric tensor gauge field strength \cite{Hull:1988jw}. The corresponding  4-form $F_{0123}$ in eleven-dimensional supergravity  is proportional to the volume form of the $dS_4$ space:
\be \label{flux}
F_{0123}\sim  \sqrt {\Lambda}\,  V_{0123} \ .
\ee
According to this model, the small value of the cosmological constant is due to the  4-form flux which has the inverse  time-scale  of the order of the age of the universe, $H_0^{-1} \sim  t_0\sim 10^{10}$ years.

Somewhat more complicated potentials are found in the case with $p=5, q=3$ with $SO(5,3)$ gauging broken down to $SO(5)\times SO(3)$ at  dS vacuum, and in the case with $p=3, q=5$ with $SO(3,5)$ gauging broken down to $SO(3)\times SO(5)$.

Near dS point all known  potentials of the gauged $N=8$ supergravity  have the following universal features \cite{Kallosh:2001gr}: the ratio ${V''\over V}$ of the eigenvalues of the mass matrix to  the potential  at the extremum   takes values $-2,4/3, 4$. This implies that
$
{m^2\over H^2_0}= \{-6 \ , 4 \ , 12\} \ .
$
In all models there is a tachyon direction with $m^2=-6H^2_0$, and for the second scalar one finds either $m^2=4H^2_0$ or $m^2=12H^2_0$, i.e.  the potential has  a saddle point. So far no  gauged  $N=8$ supergravities with 32 supersymmetries were found with dS minimum. Therefore the model with the potential (\ref{Hull}) represents also all other models, since near the critical point the tachyon has the same mass.

\subsection{Cosmology in $N=8$ models}

Since all dS critical points linked to M-theory are unstable and the  potentials are unbounded from below, one could expect that these models cannot describe the past and current evolution of the universe and play the role of the dark energy hidden sector.     However, in \cite{Kallosh:2001gr}, \cite{Linde:2001ae} it was found that if $\Lambda \sim \rho_0 \sim 10^{-120}$, the universe may stay now and during the last few billions of years  near the top of the scalar potential while the scalar field was slowly drifting away from the critical point. Here we will study this model in  detail and solve the Friedmann equations numerically.

To avoid misunderstandings, we should emphasize that  the relations $m^2=n\, H^2_0$ have been derived for dS extrema with the energy density fully supported by the scalar field potential $V(\phi)$: $m^2=n\, \Lambda/3$. In our  calculations we will {\it assume} for simplicity that this relation remains preserved  in the presence of matter fields with the energy density $\rho_M$.

In our calculations we will also  assume that initially the cold dark matter energy density was many orders of magnitude greater than the energy density of the scalar field $\phi$, $\rho_M(t_0) \gg \rho_D$. In such cases the field $\phi$ freezes very rapidly, so we can simply start our calculations assuming that its initial velocity vanished.  We   considered 4 different initial values of  the field $\phi$: \,  
$\phi_{0}=   0,\,  0.2,\, 0.3 ,\,  0.35$ in the units  $M_{Pl}=1$. The case with $\phi_{0}= 0$, where the field starts and remains at the top, corresponds to our fiducial model. The corresponding curves are shown as thick red. The blue dashed curves correspond to $\phi_{0} = 0.35$.

For each of the  values $\phi_{0}=   0,\,  0.2,\, 0.3$ we numerically found {\it different} values of $\Lambda$ such that for each of these three cases   one has   $\Omega_D=0.7$ and $H=1$ today. Here $H$ is normalized in units of the presently observed Hubble constant.
However, we could not do it for $\phi_0 = 0.35$ because the largest value of $\Omega_D$ in this case is about $0.65$; we presented this case for $H=1$ and $\Omega_D = 0.65$ for comparison.
We plot in Figure 1 the evolution of the scale factor $a(t)$, in Fig. 2 the evolution of  $\Omega_D(z)$, and in Fig. 3 the evolution of the equation of state factor $w(z)$. The point $t=0$ ($z=0$) corresponds to the present.  In all cases except $\phi_0 = 0.35$ the Big Bang occurred  at the moment $t_0 \approx -0.95 H^{-1}$ in terms of the present value of the Hubble constant $H = 1$. If one assumes, according to \cite{Bond}, that $|t_0| \approx 13.6$ billion years, one finds that each interval $\Delta t = 1$ in our figures approximately corresponds to 14 billion  years.

 \begin{figure}[h!]
\centering\leavevmode\epsfysize=5.2cm \epsfbox{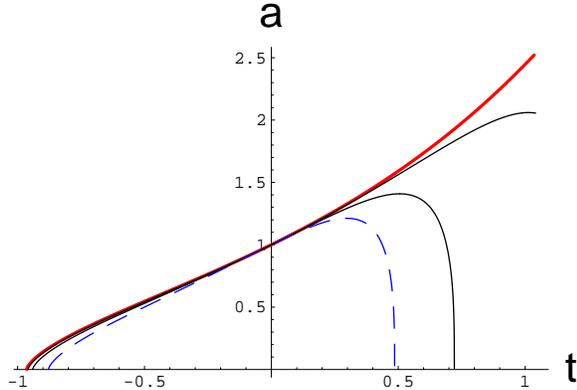}

\

\caption[fig1]
{Scale factor $a(t)$ in the model based on $N=8$ supergravity. The upper (red) curve corresponds to the model with $\phi_0 = 0$. In this case the universe can stay at the top of the effective potential for an extremely long time, until it becomes destabilized by quantum effects \cite{Kallosh:2001gr}. The curves below it correspond to $\phi_0 = 0.2$ and $\phi_0 = 0.3$. The blue dashed curve corresponds to $\phi_{0} = 0.35$.  The present moment is $t=0$. Time is given in units of $H^{-1}(t=0) \approx 14$ billion  years.}
\label{ScalefactorColl}
\end{figure}

As expected, all 3 models with $\phi_{0}=     0.2,\, 0.3 ,\,  0.35$ lead to a collapse. Still all results for the cases    $\phi_{0}= 0.2,\ 0.3 $ are compatible with observations. The case  $\phi_{0} = 0.35$  is close to be ruled out on the basis of current data with $\Omega_D=0.7$: in this model it is impossible to reach $\Omega_D=0.7$,  the largest value of $\Omega_D$ in this case is about $0.65$.  Among all of our models, this was the case with nearest   collapse in the future (7 billion  years from now). We cannot claim  that  this case is definitely  ruled out due to the cosmological observations suggesting that $\Omega_D=0.7$ today. However, the present value of $w$ for this model is very high, which is an additional evidence suggesting that this regime is disfavored by observations. For greater values of $\phi_0$ the maximal value of $\Omega_D$ becomes considerably smaller. For example, for $\phi_0 > 0.4$ the maximal value of $\Omega_D$  is  $0.56$. This means that the models with $\phi_0 > 0.4$, predicting even earlier collapse of the universe, contradict observational data.

On the other hand, the models with $\phi_{0} = 0$, $\phi_{0} = 0.2$, and $\phi_{0} = 0.3$ are consistent with the present observational data. We conclude that the $N=8$ models with the potential unbounded from below  are viable candidates for the description of the present stage of acceleration of the universe if the scalar field $\phi$ initially was in the range $0\leq \phi \lesssim 0.3$ in Planck units.

Other models of $N=8$ gauged four-dimensional supergravity  linked to M-theory \cite{Hull:1988jw} have the same important property that $m^2= -6 H_0^2$ \cite{Kallosh:2001gr}.  Therefore for the initial values of fields not far from the top of the potential, an analogous behavior is expected.

  \begin{figure}[h!]
\centering\leavevmode\epsfysize=5cm \epsfbox{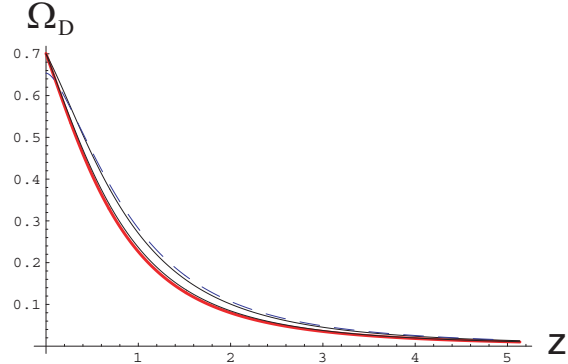}

\

\caption[fig1]
{Dark energy $ \Omega_D$ as a function of redshift $z$ for $\phi_{0}=   0,\,  0.2,\, 0.3 ,\,  0.35$. The lower (red) curve corresponds to the model with $\phi_0 = 0$. The present time corresponds to $z=0$. Note that for  $\phi_{0}=    0.35$ the maximal value of $\Omega_D$ is about 0.65.}
\label{OmegaColl}
\end{figure}

 \begin{figure}[h!]
\centering\leavevmode\epsfysize=5cm \epsfbox{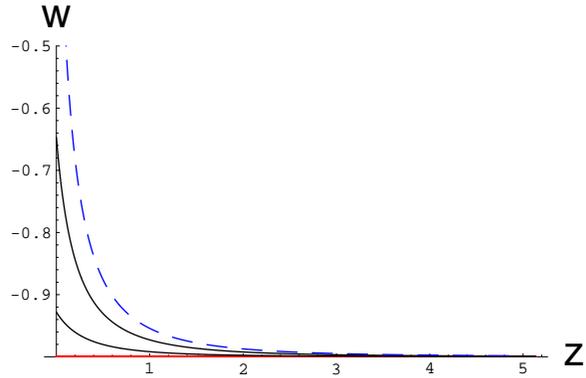}

\

\caption[fig1]
{Equation of state function $w$ as a function of redshift $z$ for $\phi_{0}=   0,\,  0.2,\, 0.3 ,\,  0.35$.  For  $\phi_{0}=    0.35$ this function sharply rises near $z=0$.  For  $\phi_{0}=    0.3$ the maximal value of $w$ is about $-0.65$. This could also seem too high, but the average value of $w$ in the important interval $z\lesssim 2$ is below $-0.9$. For  $\phi_{0}=    0.2$ the maximal value of $w$ is below $-0.9$. The red line $w =-1$ corresponds to the model with $\phi_0 = 0$.}
\label{wColl}
\end{figure}

One could argue that fixing $\Lambda \sim 10^{-120}$ and $\phi < 0.3$ in Planck units represents fine-tuning. Of course, fine-tuning of such type is present in most of the models of dark energy. But can we do better than that and avoid fine-tuning altogether?

Note that one cannot simply add a nonvanishing cosmological constant to 
N=8 supergravity keeping the curvature of the effective potential 
intact. This would violate the structure of the effective potential $V= 
\Lambda (2-  \cosh {\sqrt 2} \phi)$ and the relation $m^2 = -2\Lambda$. 
Once we increase $\Lambda$, the tachyonic mass increases as well, and 
the total lifetime of the universe becomes unacceptably small. Indeed, 
as we have found, the total lifetime of the universe in such models 
typically is $O(H^{-1}) \sim O(\Lambda^{-1/2})$. It can be shown that 
this lifetime is proportional to $\log \phi^{-1}_0$, so it can be made 
much greater than   $H^{-1}$, but only if the initial value $\phi_0$ of 
the field $\phi$ was exponentially close to zero 
\cite{Kallosh:2001gr}. When the field $\phi_0$ becomes very close to 
$\phi_0 = 0$,  the  lifetime of the universe becomes determined by  
quantum fluctuations. The total lifetime of the universe cannot exceed 
$t \sim 10^2 H_0^{-1} \sim 10^3$ billion years even for very small 
$\phi_0$ \cite{Kallosh:2001gr}.

Let us assume that for the development of life of our type the universe 
must have lifetime $t_{\rm tot} \gtrsim 14$ billion  years, which is the 
present age of the universe. This would imply that we cannot live in a 
universe with $\Lambda \gg 10^{-120}$ since such a universe will be 
short-living, with the total age much smaller than $14$ billion  years, 
unless one assumes that in the very beginning $\phi_0$ was exponentially 
small.

According to Eq. (\ref{flux}), the value of $\sqrt {\Lambda}$ is fixed by the 4-form flux, $F_{0123}\sim  \sqrt {\Lambda}\,  V_{0123}$. In general one may imagine that this flux (and, consequently, $\Lambda$), as well as the field $\phi_0$, may take different values either in different regions of the universe, or in different quantum states of the universe.  Consider all combinations of $\phi_0$ and $\Lambda$ compatible with the total lifetime of the universe $t_{\rm tot} > 14$ billion  years. If one makes an assumption that all such combinations are possible and equally probable (or equally natural), then our  investigation indicates that the  value of $\Lambda$ should very close to $\rho_0 \sim 10^{-120}$,  and the most probable time before the future collapse of the universe is $O(10)$ billion  years \cite{KallLin}. Depending on various assumptions, one finds also that the probability to have $\Omega_D$ in the interval between $0.5$ and $0.9$ at the  time $t \sim 14$ billion  years after the big bang is about $15\%$  \cite{KallLin}.  This number may change if we make different assumptions about the probability measure, but many reasonable assumptions leads to the same conclusion: Finding $\Omega_D$ in the interval between $0.5$ and $0.9$ does not require the fine-tuning that is the usual problem of many models of quintessence. This may provide a solution to the fine-tuning/coincidence problem for $\Lambda$ and $\Omega_D$ and, simultaneously, a prediction for the typical time-scale of the global collapse of the universe in the models based on $N=8$ supergravity.

\subsection{Collapsing universe in $N=4$, $N=2$  gauged supergravities}
The very first gauged supergravity with dS solutions, breaking supersymmetry spontaneously,  was found in \cite{Gates:1983ct}. The relevant part of the action of $SU(2)\times SU(2)$ gauged supergravity is

$$
g^{-1/2} L = {1\over 2} R -  {1\over 1-|W|^2}  \partial_\mu W \partial^\mu \bar W  - g^2 {1-3|W|^2\over 1-|W|^2}  \ .
$$

Here the potential depends only on $|W|$ and again
$ 
V''/V= - 2 \ ,  {m^2\over H_0^2}=-6  .
$
This means that the
cosmological evolution  in this model for initial values of the fields not far from the dS critical point at $|W|=0$ will be the same as in the previous case, explained above for $N=8$ theory.

As an example of unstable dS solutions consider here   $N=2$ $SU(2)$ gauged supergravity \cite{deWit:1984pk}. The relevant part of the action depends on the  scalar fields $R^A$ and $\bar R^A$,  $A=1,2,3$:
\begin{eqnarray}
g^{-1/2} L &=& {1\over 2} R -   \delta_{AB} \partial_\mu R^A \partial^\mu \bar R^B
 + 2g^2 [ 3 - 2\bar R^A R^B \delta_{AB} \nonumber\\ &-&  \delta_{AB}(\epsilon_{ACD} \bar R^C R^D) (\epsilon_{BEF}\bar R^E R^F)]
\nonumber \ .
 \end{eqnarray}
This action has two dS critical points. One is $SU(2)$ symmetric and $R^A_{cr}=0$, $V_{cr}= 3H_0^2= 6g^2$. The tachyonic mass is equal to $-4g^2$ and
$
{V''\over V_{cr}}= -{4\over 3}\ ,   {m^2\over H_0^2}= - 3 .
$
At the second critical point $(\bar R^A R^B \delta_{AB})_{cr}=1$, $( R^A R^B \delta_{AB})_{cr}=0$, $V_{cr}= 3H_0^2= 4g^2$. There is a tachyonic mass squared equal to $-8g^2$ and a positive mass squared $16 g^2$ and some Goldstone modes. Thus we find in the second minimum that
$ {m^2\over H_0^2}= - 6, \, 12  .
$

Thus we have shown  in examples above that despite a large variety of potentials in extended gauged supergravity, we always confirm the pattern that ${m^2\over H_0^2}$ is of the order one and there are tachyons. The development of instability and all basic features which were found by numerical solutions of the Friedmann equations in $N=8$ case will take place in other models with unstable dS solution.

\section{Gauged Supergravities with  de Sitter Attractor}

The models with dS minimum always define {\it ultimate future dS universe} since scalar fields eventually reach their constant fixed points at which the potential has a minimum, and space-time is a dS-type.  The simplest (and typical) representative of a potential in such theories (for canonically normalized fields $\phi$) is
\be
 V= \Lambda \, \cosh {\sqrt 2} \phi\ .
\ee
Here $\Lambda$ is the cosmological constant, the value of the potential at the minimum. It is related to the attractor value of the Hubble constant, $\Lambda = 3H_0^2$.  At the minimum where $V'=0$, $\phi=0$ and  $V(0)=\Lambda >0$.

\subsection{Axion-dilaton  dark energy}

$N=2$  supergravities with stable dS vacua constructed in \cite{Fre:2002pd} have special features: all moduli  are coordinates of the special K\"{a}hler and quaternionic manifold ${\cal S}{\cal T}[2,n]\bigotimes HQ[m]$. The ungauged supergravity of this type appear, e. g. in the large radius limit of superstring compactifications  \cite{Andrianopoli:1996cm}. Upon consistent gauging  of the non-compact
groups directly in $d=4$, these theories acquire a potential and become rather complicated, in general. The simplest version  with 3 vector multiplets has the axion-dilaton $S$-moduli which parametrize an $SU(1,1)\over U(1)$ part of the manifold and the so-called $T$ and $U$ moduli (Calabi-Visentini coordinates, $y_0, y_1$ ), which parametrize the ${SO(2,2)\over SO(2)\times SO(2)}$ part of the special K\"{a}hler manifold.

It was suggested in \cite{Kallosh:2002wj} to focus on the simplest part of all these models which consists of axion-dilaton only. One reason for that is the simplicity, the other reason is that axion-dilaton pair in 4d theory, coming from M/string theory, have well known global $SL(2,{\bf Z})$-symmetry. It includes the Peccei-Quinn shift of the axion. The scale of violation of this symmetry is related to the scale of the cosmological constant and mass of these fields. It is therefore interesting to study  the evolution of the universe during the last Hubble time and the future evolution towards  dS attractor and see what kind of features of dark energy and accelerated universe may be described by this model.
We will  also study  a more general model with dS future universe which include Calabi-Visentini scalars in addition to axion-dilaton fields.

 Axion-dilaton action (without a potential, before gauging) in four dimension obtained by dimensional reduction from $d=10$
  is given by
\be
{g}^{-1/2} L = {g^{\mu\nu}\partial_\mu S \partial_\nu \bar S \over (2
\rm{Im }\,S)^2} \ .
\label{AD}\ee
Here the real part of the complex modular parameter $S= {\sqrt 2}A-ie^{-{\sqrt 2} \phi}$ is the axion field $A$ and the imaginary part has the dilaton $\phi$. 

The action for canonically normalized kinetic terms 
\be
{g}^{-1/2} L = -{1\over 2}R + {1\over 2}[(\partial \phi)^2 + e^{2{\sqrt 2} \phi}(\partial A)^2]\ ,
\label{ADaction}\ee
is easily obtained either by compactification on a six-torus of the heterotic string theory in ten dimensions or by a compactification on a seven-torus from eleven dimensional M-theory supergravity action. For example, in string theory $\phi$ is related to the dilaton of the string theory and $A$ is related to the second rank antisymmetric tensor $B$-field of string theory, after a duality transformation. These are so-called fundamental dilaton and axion. In M-theory  $\phi$ is related to $g_{11, 11}$ component of the metric and $A$ is related to the third rank antisymmetric tensor field $C$. The details of $d=10$, $d=4$ connection are described in \cite{Schwarz:1992tn}.  The axion-dilaton system in string theory and ungauged supergravity  has a non-compact $SL(2,R)$ or $SL(2, {\bf Z})$ symmetry: the actions (\ref{AD}), (\ref{ADaction}) are invariant under  linear fractional transformations of the modular parameter
$
S \rightarrow  {a S +b\over c S + d}
$.
This includes the shift of the axion field by a constant,
$A \rightarrow  A + {\rm const}$.

In case of interest  a consistent  gauging   directly  in $d=4$  produces a following potential of the axion-dilaton action \cite{Fre:2002pd}, \cite{Kallosh:2002wj}:
\be \label{totpot}
 V=\Lambda \, [\cosh {\sqrt 2} (\phi-\phi_{cr})+  e^{{\sqrt 2}(\phi+\phi_{cr})}(A-A_{cr})^2]\ .
\nonumber \ee
Here $\Lambda$ is the cosmological constant, the value of the potential at the minimum. It is related to the attractor value of the Hubble constant, $\Lambda = 3H^2_0$. The attractor values of the  axion-dilaton field $S_{cr}= {\sqrt 2} A_{cr}-ie^{-{\sqrt 2} \phi_{cr}}$. At the minimum where $V'=0$,
\be
\phi=\phi_{cr} \ , \quad A=A_{cr} \ , \quad V(A_{cr}, \phi_{cr})=\Lambda >0 \ .
\ee
Near the dS vacuum solution with the axion-dilaton field fixed at the attractor values $S=S_{cr}$, the masses of the dilaton and axion fields 
are both equal to 
$
m_{\phi}^2= m_A^2= 6 \,{\Lambda\over 3}=2\Lambda  .
$

Note that the global axion shift symmetry $A\rightarrow A+\rm {const}$ is broken by the potential with the fixed value of $A_{cr}$.

The original parameters used in \cite{Fre:2002pd} are some particular combinations of gauge coupling $e_0$, Fayet-Iliopoulos term $e_1$  and magnetic rotation angles $\theta$, such that

\be
 \Lambda = e_0 e_1 \sin \theta \ ,\quad  A_{cr}= \cot \theta /{\sqrt 2} \ 
  e^{-{\sqrt 2} \phi_{cr}}= {e_0 \over   e_1 \sin \theta} \ .
\label{crit}\ee

Here the non-compact $SO(2,1)\times U(1)$ group is gauged:   $e_0$ is a gauge coupling for   $SO(2,1)$ and $e_1$ is a FI term for $U(1)$. The theory has 4 vector fields: a graviphoton and 3 vector fields for each of the $STU$ moduli. At the dS minimum the non-compact $SO(2,1)$ gauge group is broken spontaneously down to $SO(2)$: the gauge vectors associated with the non-compact generators of the $SO(2,1)$ become massive by eating one of the complex moduli.

\subsection{$S,T,U$ moduli dark energy}

An un-truncated version of the first model in \cite{Fre:2002pd} has  $S, y_0, y_1$  fields or, equivalently  $S, T, U$ fields.  We will choose a gauge where the Goldstone scalar field $y_0$ is vanishing. This will give us additional kinetic terms in the action when  $ y_1= \sigma e^{i\delta}$ are present:
\be
{2\over (1-\sigma^2)^2}[(\partial \sigma)^2 + \sigma^2 (\partial \delta)^2] \ .
\ee
The total potential is
\bea
 V =  \Lambda  \, &\Bigl[& {1+\sigma^4\over [1- \sigma^2]^2} \cosh {\sqrt 2} (\phi-\phi_{cr})\nonumber \\ &+&  {2\sigma^2\over [1- \sigma^2]^2}\sinh {\sqrt 2} (\phi-\phi_{cr})\\
&+&  e^{{\sqrt 2}(\phi+\phi_{cr})}(A-A_{cr})^2  \Bigr]\ .\nonumber
\eea
If we put the axion-dilaton fields in the minimum of the potential, i.e. at $\phi=\phi_{cr}$ and $A=A_{cr}$, the remaining action for the $\sigma$ field is
\be
{2\over (1-\sigma^2)^2}(\partial \sigma)^2 -\Lambda \,  {1+\sigma^4\over [1- \sigma^2]^2}    \ . \ee
This can be rewritten as
\be
{1\over 2}(\partial \sigma)^2 -\Lambda \, [{1+ {1\over 2}\sinh^2 \sigma]}\ ,
\ee
which gives a very simple gauged supergravity motivated model of the field with the mass-Hubble ratio
$m^2=\Lambda = 3 H_0^2$.

The model with 3 scalars, $\phi, A, \sigma$ is
\bea
&&{g}^{-1/2} L = -{1\over 2}R + {1\over 2}[(\partial \phi)^2 + e^{2{\sqrt 2} \phi}(\partial A)^2]\nonumber \\
&& ~~~ +  {2\over (1-\sigma^2)^2}(\partial \sigma)^2-\Lambda \, \Bigl[{1+\sigma^4\over [1- \sigma^2]^2} \cosh {\sqrt 2} (\phi-\phi_{cr})\label{STUaction} \\ &&~~~+   {2\sigma^2\over [1- \sigma^2]^2}\sinh {\sqrt 2} (\phi-\phi_{cr})
 +  e^{{\sqrt 2}(\phi+\phi_{cr})}(A-A_{cr})^2  \Bigr] . \nonumber
\eea

We  solved the Friedmann equations numerically for the general model with all 3 moduli fields, corresponding to the action given in eq. (\ref{STUaction}). The solutions of these equations for various initial conditions for the scalar fields show  the attractor behavior of the total solution: the scalars eventually reach their attractor values, defined by the minimum of the potential, and the universe asymptotically becomes  dS space.
Relative importance of various fields during this process depends on their initial values. However, we have found that if we begin with the values of the fields far from the minimum of the effective potential, then for a rather broad range of initial conditions the dilaton contribution becomes dominant at the late stages of the process. Therefore we will concentrate now on the simplest case of the dilaton-dominated universe.

\subsection{Generic future de Sitter universe}

Let us assume that the $\sigma$-field and the axion have already reached their fixed points, $\sigma=0$ and $A=A_{cr}$ and only the dilaton is not yet there. We also choose the vanishing value of $\phi_{cr}=0$ for simplicity:
\bea
{g}^{-1/2} L = -{1\over 2}R + {1\over 2}(\partial \phi)^2 -
\Lambda \, [ \cosh {\sqrt 2} \phi ] \ .
\label{dilaton}\eea
This model is a typical representative of $N=2$ supergravity with future dS space.

\begin{figure}[h!]
\centering\leavevmode\epsfysize=5cm \epsfbox{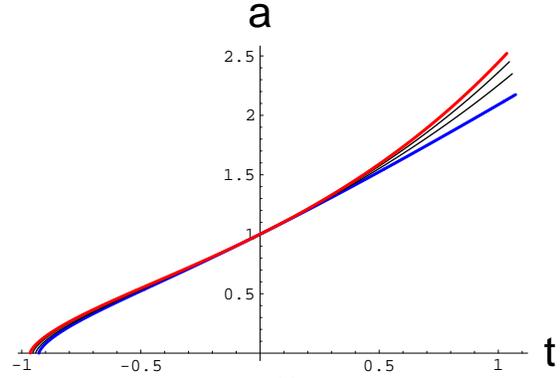}
\caption[fig1]
{Scale factor $a(t)$ in the model based on $N=2$ supergravity with a stable dS minimum for $\phi_{0}= 0,\, 0.6,\, 1.0,\, 10$. The upper (red) line corresponds to  $\phi_0 = 0$, the lower (blue) line corresponds to $\phi_0 = 10$. Blue lines in this figure and in the next two figures practically coincide with the corresponding lines for the purely exponential potential $V \sim e^{\sqrt 2 \phi}$.}
\label{sfdS}
\end{figure}

\begin{figure}[h!]
\centering\leavevmode\epsfysize=5cm \epsfbox{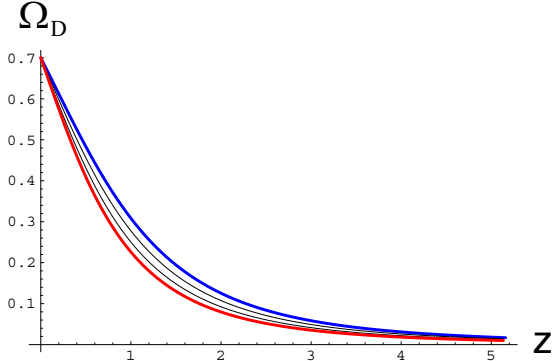}
\caption[fig1]
{Dark energy $\Omega_D(z)$ for $\phi_{0}= 0,\, 0.6,\, 1.0,\, 10$. The lower (red) line corresponds to  $\phi_0 = 0$, the upper (blue) line corresponds to $\phi_0 = 10$. }
\label{OmegadS}
\end{figure}

\begin{figure}[h!]
\centering\leavevmode\epsfysize=5cm \epsfbox{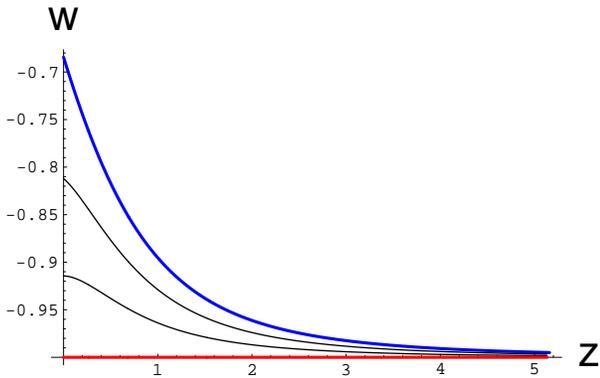}
\caption[fig1]
{Equation of state $w(z)$ for $\phi_{0}= 0,\, 0.6,\, 1.0,\, 10$. The lower (red) line corresponds to  $\phi_0 = 0$, the upper (blue) line corresponds to $\phi_0 = 10$.}
\label{wdS}
\end{figure}

We study numerical solutions of the Friedmann equations with the action given in (\ref{dilaton}). Our choice of initial conditions is $\phi_{0}= 0,\, 0.6,\, 1.0,\, 10$ and in all cases $\dot \phi_{0}=0$. As in previous models of collapsing universe, we choose for `today' $H^2=1$ and $\Omega_D=0.7$. We plot in Figure   \ref{sfdS} the evolution of the scale factor $a(t)$,  in Fig. \ref{OmegadS} the evolution of $\Omega_D(z)$ and in Fig. \ref{wdS} the evolution of the equation of state factor $w(z)$. The point $z=0$ as in collapsing models corresponds to `today'.  The red curves (the upper curve in Fig. \ref{sfdS} and the lower lines in Figs. \ref{OmegadS} and \ref{wdS}) correspond to a fiducial model where the scalar field is already at the minimum of the potential.

\

In each case we find the universe that ultimately becomes  dS space, forever expanding, never collapsing. In the past for $t\leq 0$ as well as in the future for $t>0$  for all cases we find almost the same curves for $a(t)$ and $\Omega_D(t)$. All models are very close to the fiducial model with regard to $a(t)$ and $\Omega_D(t)$.  Only the evolution of the equation of state factor $w(t)$ varies from case to case. The importance of this fact is that it may at least in principle become observable. For all models with
$\phi_{0}> 0.6$ we find $\dot w>0$ at $t=0$, for $\phi_{0}=  0.6$ $\dot w=0$ at $t=0$ and for $\phi_{0}<0.6$ $\dot w<0$ at $t=0$.

We should note that this model requires fine tuning of initial conditions. We began the calculations at some $\phi = \phi_0$, and adjusted the value of the effective potential in its minimum in order to obtain $\Omega_D = 0.7$ at the time $t_0 \approx 14$ billion  years. However, this is the standard fine-tuning that is present in most of the models of quintessence; see a discussion below. In this respect it is interesting that the models with dS {\it maximum} considered in the previous section may not require  fine-tuning \cite{KallLin}.

\

\section{M-theory and Dark Energy with Exponential Potentials}

The results obtained above are closely related to investigation of dark energy in the theories with purely exponential potentials. Indeed,  the blue line corresponding to the initial condition $\phi_0= 10$ (the lower line in Fig. \ref{sfdS} and the upper line in Figs. \ref{OmegadS} and \ref{wdS})   practically does not change when   $\phi_0$ increases even further. This result has an interesting interpretation. In the limit $\phi_0 \gg 1$  (for $\sigma=0$ and $A=A_{cr}$) the dilaton potential (\ref{totpot}) can be represented as a simple exponent:
\be
V= \Lambda e^{\sqrt{2}\phi} \ .
\ee
In this case increase of $\phi_0$ to $\phi_0'$ can be absorbed by rescaling  $\Lambda \to \Lambda \exp{\sqrt 2 (\phi_0- \phi_0')}$. This is what we are doing  in our calculations when we are normalizing the present Hubble constant $H$ to $H=1$. As a result, for all $\phi_0$ in the exponential potential one has the same curves $\Omega(z)$ and $w(z)$. These curves practically coincide with the  blue lines in Figs. \ref{OmegadS} and \ref{wdS}.

This conclusion has important implications. The first models of dark energy were based on investigation of the theories with exponential potentials   \cite{Wett}. For $V= \Lambda e^{\lambda \phi}$  different possibilities were studied, depending on the value of $\lambda$. In the presence of cold dark matter with equation of state $p=0$, the interesting regions are:
\begin{itemize}
\item For  $\lambda^2<3$  the system has a {\it late-time attractor} with fixed point values  $\Omega_D^{attr}=1$ and $w^{attr}= \lambda^2/3-1$.
\item
For  $\lambda^2>3$  there is a stable solution of a  scaling type, a {\it global attractor} with fixed values  $\Omega_D^{attr}=3/\lambda^2$ and $w^{attr}=0$.
\end{itemize}
In the first case one does not have $\Omega_D=0.7$ in the attractor regime.
In the second case the power-low regime with $\Omega_D \approx 0.7$  is possible. However, since $\Omega_D$ is approximately constant,
it is difficult to reconcile this regime with the big bang nucleosynthesis which requires that at the stage of the nucleosynthesis $\Omega_D$ must be very small \cite{Wett}. Moreover, the limit $w \to 0$ is not good for quintessence. Therefore the general lore is that the theories with exponential  potentials cannot describe dark energy.

In our paper (see also \cite{WellAl,LopesFranca:2002ek}) we study a  regime that occurs {\it before  the late time attractor is reached}, for $\lambda^2<3$. We simply  check whether we can find such parameters of the theory that would allow us to consistently describe the observational data. And the answer to this question is positive. If one chooses proper initial conditions (the universe  dominated by CMB), initial values of $\Omega_D$ become negligible, so one has no problems with  the big bang nucleosynthesis. Meanwhile at some later stage $\Omega_D$ can grow and reach the desirable value  $0.7$. Therefore exponential potentials can describe the present stage of acceleration of the universe. This description requires some fine-tuning, but it is not worse than it is in many other models of dark energy. Our conclusion agrees with the one  obtained  in \cite{LopesFranca:2002ek}.

Another problem associated with exponential potentials is that they must be sufficiently flat. It is well known that the universe, in the absence of ordinary matter, will accelerate only for $\lambda^2 <  2$; for $\lambda^2 =  2$ one would have a linear regime $a \sim t$ with $\ddot a = 0$. However, in order to explain observational data we do not have to assume that the universe will keep accelerating in the distant future, so the requirement $\lambda <\sqrt 2$ is not necessary. On the other hand, our  numerical studies show that for $\lambda > 2.2$ the value of $\Omega_D$ never reaches $0.7$. For $\lambda > 1.7$ one can have $\Omega_D = 0.7$, but one has $w > -0.6$ at present, which is perhaps too high \cite{LopesFranca:2002ek}.

We may conclude therefore that for $\lambda^2<3$,  when the system has a {\it late-time attractor}, the reasonable description  of the present and past can be easily achieved at the stage {\it before the late attractor is reached}. Eventually, in the long term future the  fixed point values  will be reached, $\Omega_D \rightarrow 1$ and $w \rightarrow \lambda^2/3-1$.

Recently it was pointed out that it might be difficult to describe eternally accelerating universe with $\lambda^2 <  2$ in terms of M/string theory because it has event horizon \cite{Hellerman:2001yi}. In   case $\lambda^2 =  2$ the universe does not accelerate in long term future and has no event horizon, so this problem does not appear. Moreover, if one takes into account ordinary matter, then the universe at the asymptotically large time slowly decelerates. But one can easily describe by this model a universe  that presently accelerates and has $\Omega_D = 0.7$. The present value of $w$ in this case is about $-0.7$, but the average value  $\bar w$ (\ref{barw}) in this case is $-0.79$, which looks pretty safe from the point of view of the present observational data.

It has been observed in \cite{Hellerman:2001yi} that for the positive exponential potentials related to higher dimensional supergravities the typical value is $\lambda^2\geq  6$. However this constraint may be decreased due to the presence of the FI terms, as shown in   \cite{Townsend:2001ea}.  An interesting question  to be explored  is  the M/string theory origin of  potentials with $\lambda^2\leq  2$ which could provide us with cosmological models of the presently accelerating universe.

In our case the model with $V= \Lambda e^{\sqrt{2}\phi}$ appears as a special limiting case of the $N=2$ model considered in the previous section. But a purely exponential $V= \Lambda e^{\sqrt{2}\phi}$ appears in some other theories as well.

Indeed, a model of $N=2$,  $d=4$ gauged supergravity with 3 vector multiplets, without hypermultiplets,  described in \cite{Andrianopoli:1996cm},  has exactly such a potential:
\be
g^{-1/2} L = -{1\over 2} R +  {1\over 2}  \partial_\mu \phi \partial_\nu \phi \, g^{\mu\nu} -{\xi^2\over 2} e^{\sqrt {2} \phi} \ ,
\label{sugraEXP} \ee
where the other few scalar fields, which are coordinates of ${SU(1,1)\over U(1)}\times {SO(2,2)\over SO(2)\times SO(2)}$  coset space are constant. This Lagrangian  can be deduced from the complete action given in \cite{Andrianopoli:1996cm}. The relevant symplectic sections are $X^\Lambda, F_{\Lambda}= \eta_{\Lambda \Sigma} S X^\Lambda; X^\Lambda X^\Sigma \eta_{\Lambda \Sigma}=0,
\eta_{\Lambda \Sigma}=(1, 1, -1, -1)$. The gravity-dilaton part of the action is
\be
-{1\over 2} R + {\partial_\mu S \partial^\mu \bar S \over (2 \rm{Im }\, S)^2} + { \xi^2 \over  2\,  \rm{Im }\, S} \ .
\label{dil}\ee
The complete gauged supergravity action is consistent under condition that $\rm{Im }\, S <0$ which provides a positivity condition for the kinetic terms of scalars and vectors.
For $\rm{Im }\, S= -e^{-\sqrt 2 \phi}$ the action (\ref{dil}) is reduced to (\ref{sugraEXP}) with a positive definite potential.

The potential is defined in eqs. (9.54) and (9.58) of \cite{Andrianopoli:1996cm}.
The source of such positive potential is due to Fayet-Iliopoulos term $P_\Lambda^x$ with $(P_4^x)^2= \xi^2$. This  is possible for an Abelian gauge group. For the positive sign the  Abelian gauge group with FI terms must be neither the graviphoton, nor the gravidilaton, but one of the remaining vectors in the direction where $\eta_{44}=-1$.

The action (\ref{sugraEXP}) can be also identified with the model of Ref. \cite{Fre:2002pd} under condition that there is no gauging of the $SO(2,1)$ group, $e_0=0$,   the rotation angle is vanishing,  $\theta=0$, and only FI terms with $e_1$ are present. In such case this model is reduced to the model proposed in
\cite{Andrianopoli:1996cm} and described by Eqs. (\ref{totpot}), (\ref{crit}). The contribution to the potential surviving the limit $e_0\rightarrow 0$ and $\theta \rightarrow 0$ is coming from the term $e^{\sqrt {2} \phi}(\Lambda e^{\sqrt {2} \phi_{cr}}A_{cr}^2)\rightarrow  {e_1^2\over 2} e^{\sqrt {2} \phi} $. Since $e_1$ is a FI term $\xi$ this clarifies the origin of the positive exponential potential in $N=2$ supergravities.

\section{$N=1$ Supergravity}

\subsection{Supergravity  quintessence model}

Now we take lessons learned in extended $N\geq 2$ supergravities and use them as a guide for $N=1$ models which may give us a successful description of the recent past of our universe, including the acceleration period, and lead to particular prediction about the future.

The first example is the one already known as a `supergravity quintessence' model \cite{Brax:2001ah} with the potential:
\be
V= e^{\phi^2\over 2}{M^{4+\alpha}\over \phi^\alpha} \ .
\label{BraxPot}\ee
This model has a dS minimum at $\phi_{cr}= \sqrt \alpha$. Interestingly,  the mass of the scalar field near dS minimum is exactly the same as in the $N=2$   model  discussed above:
\be
m^2= 6H_0^2 \ .
\label{6}\ee

We solved the Friedmann equations for the cases with $\alpha= 4, 6, 11$. In case of initial values of the fields not far from the dS attractor value, all features of the solution ($a(t), \Omega_D (t), w(t)$) are practically the same as  in our $N=2$ future dS model. However, if one considers the model with $\alpha =11$ \cite{Brax:2001ah} and take the initial value of the field far away from the minimum of the effective potential at $\phi = \sqrt 11$, which is necessary to approach the tracker solution, then $w(t)$ rapidly increases from $-0.8$  and for a long time stays above $-0.2$.  The average value  $\bar w$ (\ref{barw}) in this case is about $-0.65$. This makes this scenario rather vulnerable \cite{Copeland}.

 \begin{figure}[h!]
\centering\leavevmode\epsfysize=4.7cm \epsfbox{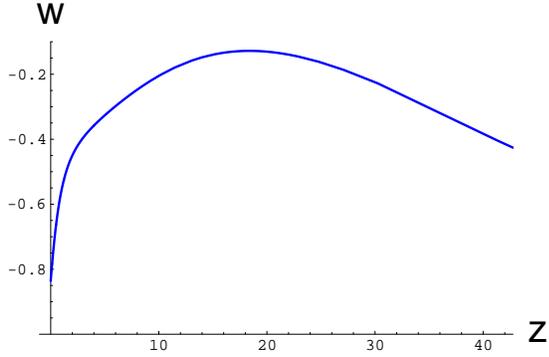}

\

\caption[fig1]
{The values of $w$ for $\phi_0 = 1$ in the supergravity quintessence model of Ref. \cite{Brax:2001ah} as a function of redshift $z$. The value of $w$ at present is about $-0.8$, in agreement with \cite{Brax:2001ah}. But  at $z > 1$ the value of $w$ is very high.}
\label{Brax}
\end{figure}

A complete supergravity model, underlying the potential (\ref{BraxPot}) is  rather complicated and fine-tuned. In addition to the quintessence field, it  involves  several different fields charged under an abelian group, $X$-fields of charge 1, $Y$ fields of charge -2.   Various additional assumptions about the non-minimal K\"{a}hler potential and   superpotential are necessary to finally come up with the everywhere positive supergravity  quintessence potential (\ref{BraxPot}).

One could argue that this fine-tuning of a supergravity model is a reasonable price for the tracker nature of the solution. However,  when one modifies the inverse power law potential $ {M^{4+\alpha}\over \phi^\alpha}$ by multiplying it by an $e^{\phi^2\over 2}$ as in Eq. (\ref{BraxPot}), the potential acquires a minimum. All of the  solutions at the time when  $\Omega_D$ becomes equal to $0.7$ enter the immediate vicinity of this minimum. The value of the potential in this minimum needs to be fine-tuned if we want this moment to coincide with the present time.

Thus, this model requires one fine tuning, exactly as all other models with de Sitter minimum or with the exponential potential considered in our paper. In this sense, tracking in the model of \cite{Brax:2001ah} does not seem to provide additional benefits because it does not resolve the fine-tuning problem. That is why we would like to look for simpler models  based on $N=1$ supergravity that may also describe dark energy.

\subsection{Generic $N=1$ supergravity and Pol\'{o}nyi model}

The general potential of $N=1$ supergravity consists of an $F$-term and a $D$-term. Here for simplicity we will restrict ourselves to the study of the simplest models without a $D$-term. In this case
\begin{equation}
 V(z)
 = e^{\cal K}\left[ -3WW^*+({\cal D}^iW)g^{-1}
 {}_i{}^j({\cal D}_jW^*)\right].
\label{Vtotal}
\end{equation}
Here ${\cal K}$ is the {K\"ahler} potential and $W(z_i)$ is a superpotential of a chiral superfield $z_i$. The covariant derivative on $W$ is
$
  {\cal D}^iW= \partial  ^i W+\left( \partial ^i {\cal K}\right) W\ .
\label{DiW}
$
As a simplest example, consider the model with the minimal {K\"ahler} potential ${\cal K} = zz^*$ and the linear superpotential $W(z) = \mu^2(z+\beta)$. This is the famous Pol\'{o}nyi model, which serves as a standard part of models with gravity mediated SUSY breaking, see e.g. \cite{Polonyi:1977pj}.
It is convenient to represent the complex field $z$ as a sum of two canonically normalized fields, $z = (\phi+i\chi)/\sqrt 2$. One can show that the minimum of the effective potential occurs at $\chi = 0$, so we can restrict ourselves to investigation of $V(\phi)$, where
\begin{eqnarray}
 V(\phi)
 &=& {\mu^4}\ e^{\phi^2\over 2M_p^2}\Bigl[\Bigl(1+{\phi\over \sqrt2 M_p}\Bigl({\phi\over \sqrt2 M_p} +\beta\Bigr)\Bigr)^2\nonumber \\ &-&   3\Bigl({\phi\over \sqrt2 M_p}+\beta\Bigr)^2\Bigr].
\label{Vtotal2}
\end{eqnarray}
Here we temporarily restore $M_p$ in our equations. Note that in the limit $M_p\to \infty$ corresponding to global SUSY this potential becomes exactly flat, i.e. the effective mass of the moduli field $\phi$ vanishes. For $\beta = O(1)$ this potential has a minimum at $\phi = O(M_p)$, and the mass $m$ of the field $\phi$ in the minimum is $O(\mu^2/M_p)$.

Usually it is assumed that the Pol\'{o}nyi field leads to supersymmetry breaking in the observable sector. It is supposed to have  mass $m \sim 10^3$ GeV. Since this mass appears only due to SUSY breaking because of the  gravitational effects, it is protected with respect to radiative corrections. As one can easily see from Eq. (\ref{Vtotal2}), the value of $H^2 \sim V(\phi)/M_p^2$  for $|\beta| \ll 1$  and $|\phi|\ll \sqrt 2 M_p)$ is of the same order of magnitude as  $ m^2(\phi) =  V''$. This is a rather general (though not unavoidable) property  of moduli fields in $N=1$ supergravity: Typically one has $ H^2  \sim  V(\phi) /M_p^2 \sim |m^2(\phi)|$, see \cite{Dine:1983ys} and references therein. This is very similar to the situation in extended supergravity described above. However, in $N=1$ supergravity this relation is not rigidly fixed. In particular, one can fine-tune the parameter $\beta$ to be equal to $2-\sqrt 3$. In this case the effective potential vanishes in its minimum, which is necessary to avoid having an enormously large cosmological constant $\Lambda \sim \mu^4$ and still have supersymmetry breaking with $m_{3/2} \sim m \sim 10^3$ GeV.

\begin{figure}[h!]
\centering\leavevmode\epsfysize=5cm \epsfbox{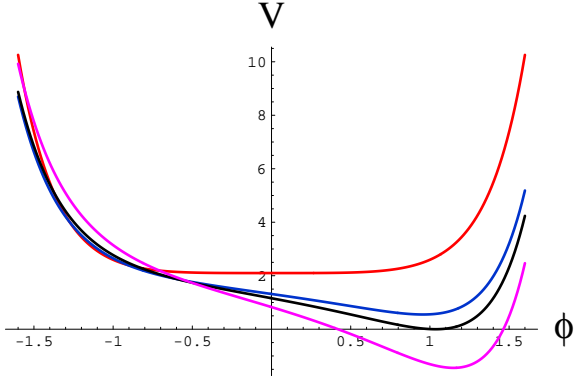}

\

\caption[fig1]
{Pol\'{o}nyi field potential for $\beta = 0$ (symmetric potential with a minimum at $V>0$), $\beta = 0.2$ (minimum at $V>0$), $\beta = 2-\sqrt 3$ (minimum at $V=0$), and $\beta = 0.4$ (minimum at $V<0$). In all of these cases $|V''| = O(|V|)$ for $|\phi| \lesssim 1$. The potential is shown in units of $m^2$; the field is shown in units of $M_p$. Potentials with $\beta <0$ can be obtained by the change $\phi \to -\phi$. }
\label{Polpot}
\end{figure}

However, if one can achieve supersymmetry breaking with vanishing cosmological constant by a different method, then instead of the field with   mass $m \sim 10^3$ GeV (or in addition to it)  one can consider an ultra-light Pol\'{o}nyi field $\phi$ with  $m \sim 10^{-33}$ eV as a part of the dark energy hidden sector. Then the absolute value of the effective potential in its minimum for $\beta = O(1)$ is $O(10^{-120})\sim \rho_0$, so there is no reason to fine-tune $\beta$ to be $2-\sqrt 3 \approx 0.268$, see Fig. \ref{Polpot}. It is important, that for $V(\phi) > 0$ and $\beta = O(1)$ one has a generic relation $ H^2  \sim   m^2(\phi) $, just like in extended supergravity. As a result, many of our results obtained in application to the $N=8$ and $N=2$ models can be easily extended to a large class of $N=1$ supergravity models.

\begin{figure}[h!]
\centering\leavevmode\epsfysize=5cm \epsfbox{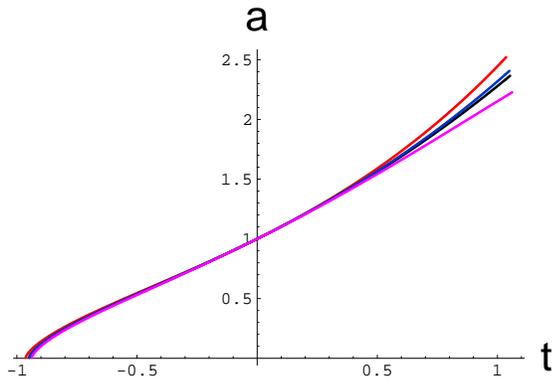}

\

\caption[fig1]
{The scale factor $a(t)$ for the Pol\'{o}nyi field potential. At large $t$ the upper curve corresponds to  $\beta = 0$, the next one - to $\beta = 0.2$, then $\beta = 2-\sqrt 3$, and $\beta = 0.4$.  }
\label{Pola}
\end{figure}

\begin{figure}[h!]
\centering\leavevmode\epsfysize=5cm \epsfbox{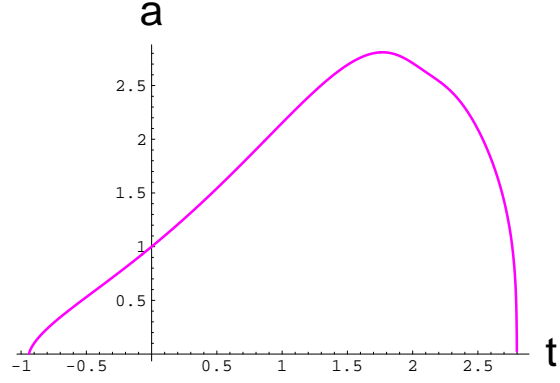}

\

\caption[fig1]
{The future of the universe in the Pol\'{o}nyi model for $\beta = 0.4$ and $\phi_0 = -1$. The present time corresponds to $t=0$. The universe in this regime would accelerate for the next 20 billion years, but then eventually collapse. In the models with $\beta > 0.4$ the collapse occurs much earlier.}
\label{Polfuture}
\end{figure}

\begin{figure}[h!]
\centering\leavevmode\epsfysize=4.5cm \epsfbox{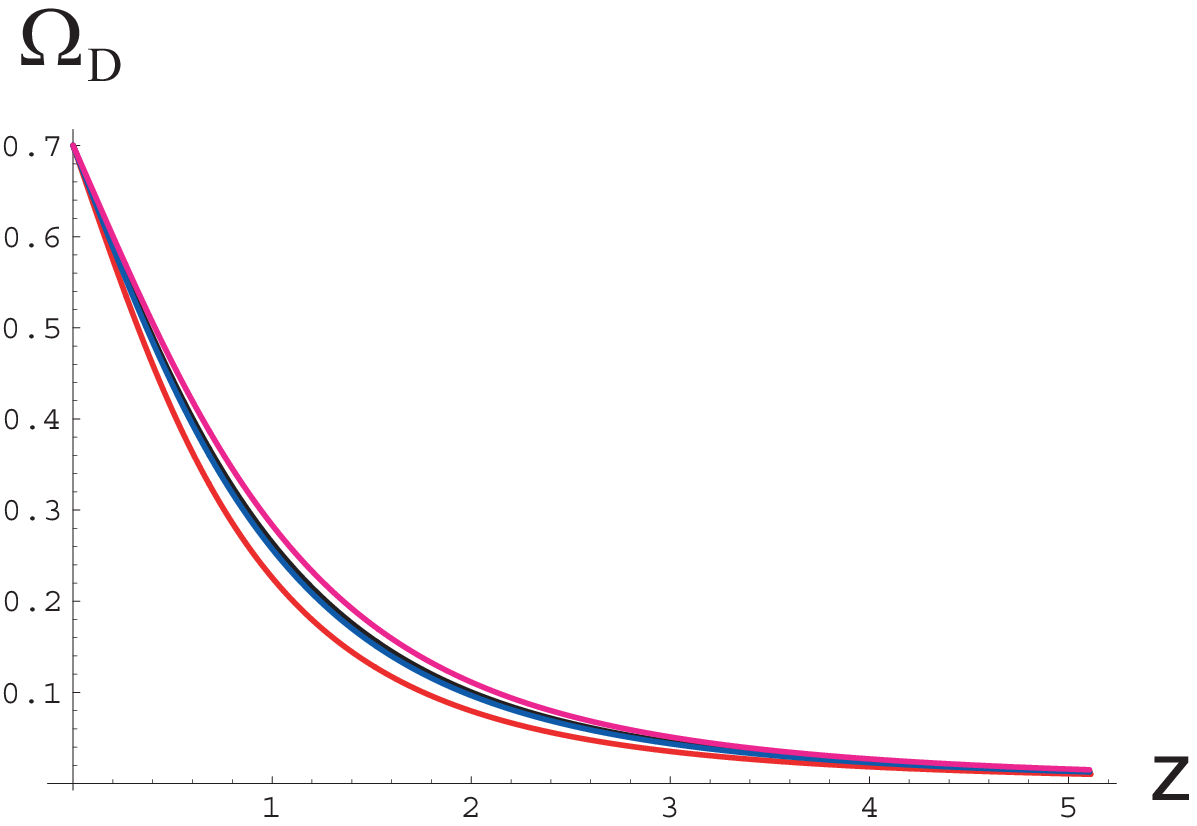}

\

\caption[fig1]
{The values of $\Omega_D$ for the Pol\'{o}nyi field potential  as a function of redshift $z$. The lower (horizontal) curve corresponds to  $\beta = 0$, the next one - to $\beta = 0.2$, then $\beta = 2-\sqrt 3$, and $\beta = 0.4$.  }
\label{Polom}
\end{figure}

\begin{figure}[h!]
\centering\leavevmode\epsfysize=4.8cm \epsfbox{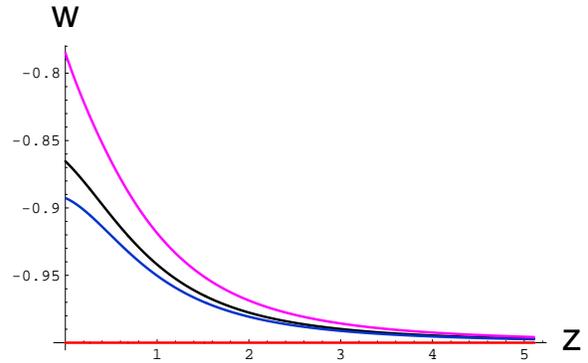}

\

\caption[fig1]
{The values of $w$ for the Pol\'{o}nyi field potential as a function of redshift $z$. The lower (horizontal) curve corresponds to  $\beta = 0$, the next one - to $\beta = 0.2$, then $\beta = 2-\sqrt 3$, and $\beta = 0.4$.  }
\label{Polw}
\end{figure}

In particular, Pol\'{o}nyi model with $|\beta| < 2-\sqrt 3 \approx 0.268$  leads to asymptotically de Sitter universe, just like in $N=2$ model of Ref. \cite{Fre:2002pd}. The fine-tuned model  with $|\beta| = 2-\sqrt 3$ asymptotically leads to Minkowski space. Meanwhile all models with $2-\sqrt 3 < |\beta| $ lead to a collapsing universe. However, just like the $N=8$ model, the $N=1$ models with $2-\sqrt 3 < |\beta| \lesssim 0.5 $ can describe dark energy in an accelerating universe, see Figs. \ref{Pola} - \ref{Polw}. These figures show the results of calculations where we for definiteness took the initial value of the field $\phi$ equal to $\phi_0 = -1$.

Thus, even the  simplest models based on $N = 1$ supergravity with $m \sim 10^{-33}$ eV can provide a natural description of dark energy because of the generic relation $|m^2| \sim H^2$ that is satisfied in many of such models. 

\subsection{The axion dark energy model}

One of the popular models of quintessence based on N=1 supergravity is the axion model.
The original axion model of  quintessence was proposed in \cite{Frieman:1995pm}. It  has the  potential: 
\be
V =\Lambda [\cos \left(\phi/f\right)+C] \ . 
\ee
In \cite{Frieman:1995pm} and in most of the consequent studies of the accelerated universe in this model it was assumed  that $C=1$. This means that  the cosmological constant was assumed to vanish in the minimum of the potential,  just as in many  other quintessence models.
The positive definiteness of the potential  with $C=1$ and the fact that it has a minimum at $V=0$ could be motivated, in particular, by global supersymmetry arguments. In supergravity and M/string theory   the argument from global supersymmetry   requiring  $C=1$ is not  longer valid. The derivation  of the  value of the  constant term $C$ in the  axion  potential from any fundamental theory is not available. 

In \cite{Choi:1999xn} the axion model of quintessence was studied using the arguments from M/string theory. These arguments were based on the properties of the membrane instantons.  Some assumptions were made concerning the superpotential and K\"ahler potential in the effective  $N=1$ supergravity potential (\ref{Vtotal}) motivated by M/string theory. The axion potential was presented there in the form
\be
V= \Lambda \cos \left(\phi/f\right)\ ,
\label{axion}\ee 
without any constant part. Perhaps it is possible to add to this potential a positive constant $C$, but here we will analyse this model without making any such  additions. This  potential has a maximum at $\phi = 0$, $V(0)=\Lambda$, and a minimum at $\phi = f\pi$, $V(f\pi)=-\Lambda$. Therefore the universe collapses when the field $\phi$  rolls down from  the top of this potential.
The  axion model with the potential (\ref{axion}) given  in \cite{Choi:1999xn}  has the properties that are qualitatively similar to those described in our paper if one assumes that $f = O(M_p)$. 
Indeed, the curvature of the effective potential in its maximum at $\phi = 0$ is given by
\be
m^2= -{\Lambda\over f^2} = -{3\over f^2} H_0^2 \ .
\label{axionpot}\ee 
If one takes, for definiteness, $f = M_p = 1$, one finds
\be
m^2= -{\Lambda} = -{3} H_0^2 \ ,
\label{axionpot2}\ee 
and for $f = M_p/\sqrt 2$ one has
\be
m^2= -{2\Lambda} = -{6} H_0^2\ ,
\label{axionpot3}\ee 
exactly as in the $N=8$ model.

\begin{figure}[h!]
\centering\leavevmode\epsfysize=4.8cm \epsfbox{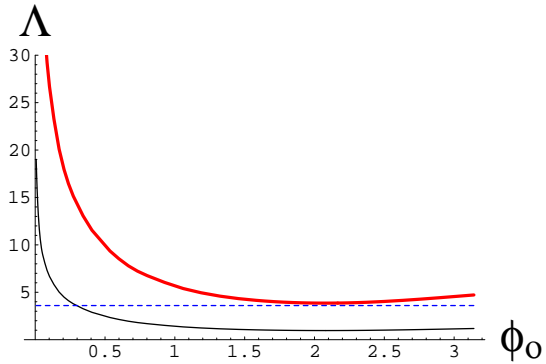}

\

\caption[fig1]
{The region below the thick (red) line contains all possible $\Lambda$ and $\phi_0$ corresponding to the total lifetime of the universe greater than $14$ billion years as predicted by the axion model (\ref{axion}). The (blue) dashed line $\Lambda \approx 3.6 \rho_0$ separates this region  into two equal area parts.  The region below the thin (black) curve corresponds to all universes with the lifetime greater than 28 billion years, i.e. to the universes that would live longer than $14$ billion years after the present moment. The area under this curve is 4 times smaller than the area under the thin (red) curve.}
\label{axfig}
\end{figure}

It is quite possible to describe dark energy using this model. The results of the investigation  of this model are very similar to our results for $N=8$ supergravity, so we will not reproduce here the figures for $a(t)$, $\Omega_D$ and $w$. Instead  we present in Figure \ref{axfig} the range of all possible values of $\Lambda$ and $\phi_0$ which allow the universe (described by the model with $f = M_p$)  to live longer than 14 billion years (the area under the thick (red) curve). $\Lambda$ is given in units of $\rho_0$ and  $\phi$  in units of $M_p$. Note that  this graph is periodic and symmetric under reflection $\phi_0 \to -\phi_0$. Therefore it is sufficient to present our results for $0<\phi_0<f \pi$.

To conclude, the study of extended supergravities attracted our attention to the properties of  other  realistic models, like the axion one. As we see from the Figure \ref{axfig}, the set of parameters that allow the universe to have the total lifetime of  more than 28 billion years (i.e. to live another 14 billion years from now on) is 4 times smaller than the set of parameters that allow our universe to live only 14 billion years. This implies that the axion model with  $V= \Lambda \cos \left(\phi/f\right)$ and $f = O(M_p)$ can describe dark energy, but as in many previous cases, the universe can be expected to collapse within the next 10 - 20 billion years.  The future collapse allows the anthropic considerations to be applied to such models, which in turn leads to a reasonable order of magnitude prediction of the value of the cosmological constant \cite{KallLin}.

\section{Discussion}

In this paper we studied various possibilities to describe dark energy in supergravities and the future of the universe in such models. Most of the 
previous works on this subject  concentrated on the attempts to obtain  inverse power law tracker potentials  in $N=1$ supergravity \cite{Binetruy:1998rz,Brax:2001ah}. These models required introduction of rather complicated superpotentials and K\"{a}hler potentials of several superfields and many assumptions.

 In this paper we have chosen a different strategy. First of all, we considered several toy models based on extended supergravity which may have closer relation the M-theory. In addition we considered  the simplest dark energy models in $N=1$ supergravity. All of these models share a very interesting feature: The absolute value of the effective mass squared of the scalar field responsible for the dark energy of the universe  is of the same order as the effective potential of this field, which implies that $|m^2| = O(H^2)$. Whereas in the phenomenological models of quintessence this property usually is {\it required} for their consistency, in supergravity this feature is rather common and sometimes it is even unavoidable. In particular, in all known versions of extended supergravity with de Sitter solutions with the Hubble constant H, scalar field masses are always quantized: $m^2 = n\, H^2_0$, where $n$ are some integers which can be either positive or negative \cite{Kallosh:2001gr,Fre:2002pd,Kallosh:2002wj}. For example, in all versions of $N=8$ supergravity of this type one always has a tachyon field with mass $m^2 = V''_0 = -6 H^2_0 = -2V_0$, where $V_0$ is the value of the effective potential in its extremum corresponding to de Sitter solution, and $V''_0$ is the curvature of the potential at that point \cite{Kallosh:2001gr,Kallosh:2002wj}.
This makes all models with $|m^2| = O(V)$ interesting candidates for the role of the dark energy.

In all models with de Sitter vacuum state with $m^2 \sim V_0 > 0$ the field $\phi$ slowly rolls to the minimum of the effective potential and the universe eventually reaches de Sitter state with $H^2_0 = V_0/3$.

On the other hand,  all $N=8 ,\,  N=4 ,\,  N=2$ models with $m^2 = V''_0 = -6 H^2_0 = -2V_0$, as well as  $N=1$ Pol\'{o}nyi models with $\beta >2-\sqrt 3$,  describing the present state with $\Omega_D \approx 0.7$, lead to the following generic prediction concerning the future of the universe: Our universe is going to collapse within the time comparable to its present age $t \sim 14$ billion years. Similar results are valid for the axion model of  quintessence with the potential $V =\Lambda [\cos \left(\phi/f\right)+C]$ for $C<1$.

The possibility that a flat universe may collapse in the future was known for quite a while \cite{MTW} -\cite{Heard:2002dr}. However, this possibility seemed to be rather extravagant, especially in view of the observations suggesting that the universe is accelerating. And even though we knew that the universe may collapse  in a distant future, there was no reason to expect that this may happen relatively soon.

Now the situation becomes quite different. Among all models of dark energy based on extended supergravity only the $N=2$ model of Ref. \cite{Fre:2002pd} leads to a stable de Sitter space in the future; all other models lead to a collapse. Among the simplest $N=1$ Pol\'{o}nyi models only the models with $\beta < 0.268$ lead to a stable de Sitter space. Similarly, among all axion models with the potential $V =\Lambda [\cos \left(\phi/f\right)+C]$ only the models with $C \geq 1$ lead to a stable dS space. All other models predict that our universe should collapse within the next $O(10)$ billion years.

Of course, all of the models considered in our paper are just toy models. We assumed that the dark energy hidden sector can be successfully incorporated into the theory of elementary particles and that the cosmological constant problem in the observable sector can somehow be solved. But this is a general issue with all models of dark energy. On the other hand,   the new class of models may provide an unusual solution to the coincidence problem.

Indeed, the total lifetime of the universe in $N=8$ theories with de Sitter solutions \cite{Hull:1988jw,Kallosh:2001gr,Kallosh:2002wj}  is $O(H^{-1}) \sim O(\Lambda^{-1/2})$. This lifetime can be few times greater, but only if the initial value $\phi_0$ of the field $\phi$ is exponentially close to $\phi=0$. Thus, large values of $\Lambda$ are forbidden since they do not allow the long-living universes. 

If one considers all combinations of $\phi_0$ and $\Lambda$ compatible with the total lifetime of the universe $t_{\rm tot} > 14$ billion  years and  assume that all such combinations are equally probable,  one finds that the  value of $\Lambda$ should very close to $\rho_0 \sim 10^{-120}$,  and the most probable time before the future collapse of the universe is $O(10)$ billion  years.   This may provide a solution to the fine-tuning/coincidence problem for $\Lambda$ and $\Omega_D$  simultaneously predicting  the typical time-scale of the global collapse of the universe in the models based on $N=8$ supergravity  \cite{KallLin}. Similar conclusion is valid for the simplest $N=1$ Pol\'{o}nyi models with $\beta > 0.268$.

Interestingly, all models that lead to the ``doomsday prediction'' have some  features that may allow us either to rule them out or to estimate the time remaining until the global collapse. For example, the $N=8$ model with $\phi_0 < 0.3$ and $\Omega_D = 0.7$ would lead to a collapse within the next 10 billion years. Meanwhile, the model with $\phi_0 = 0.35$ would lead to a collapse within the next 7 billion years, see Fig. \ref{ScalefactorColl}. However, this model leads to a maximal value of $\Omega_D \approx 0.65$, which is possible but not particularly favoured by the recent observations \cite{Bond}. The models with $\phi_0 > 0.4$  would place the doomsday  even much closer to the present moment, but they lead to $\Omega_D < 0.56$, which is incompatible with the present cosmological measurements of $\Omega_D$.

An additional information can be obtained by the measurements of $w$. In all models predicting the doomsday in the near future, the value of $w$ tends to grow significantly at small $z$, i.e. at the present epoch, see Figs. \ref{wColl}, \ref{Polw}. It is difficult to determine the time dependence of $w$ using CMB experiments alone \cite{Bond}, but one can combine them with the supernova observations,  counts of galaxies and of clusters of galaxies, and with investigation of weak gravitational lensing \cite{WellAl,Copeland,Weller:2001gf}. Even in this case it is difficult to find  the equation of state of dark energy $w(z)$. However, if one takes the cosmological models based on supergravity seriously and realizes that our future is at stake, one gets an additional strong incentive to develop observational and theoretical cosmology. It was never easy to look into the future, but it is possible to do so and we should not miss our chance.

\

It is a pleasure to thank A.~Albrecht, T.~Banks, T.~Dent, M.~Dine, G.~Felder, J.~ Frieman, N. Kaloper, A. Klypin, L. Kofman, D. Lyth, S.~Perlmutter, A.~ Starobinsky, L. Susskind, P. Townsend, A. Vilenkin and C.~Wetterich for useful discussions. This work was supported by NSF grant PHY-9870115. The work by A.L. was also supported
by the Templeton Foundation grant No. 938-COS273.  The work of S.P. was also supported by Stanford Graduate Fellowship foundation. M.S. work was also supported by DOE grant number DE-AC03-76SF00515.

\end{document}